\documentclass[runningheads]{llncs}
\usepackage{graphicx}
\usepackage{hyperref}

\usepackage{booktabs}
\usepackage{amsmath}
\usepackage{amsfonts}
\usepackage[utf8]{inputenc}

\usepackage{subcaption}
\usepackage{enumitem}
\usepackage{wrapfig}
\usepackage{multirow}
\usepackage{rotating}
\usepackage{diagbox}
\usepackage{ntheorem}

\usepackage{floatrow}
\begin{document}
\title{Identifying DNS-tunneled traffic with predictive models}
%
%
\author{Andreas Berg\inst{1}\orcidID{0000-0001-5932-3304} \and
Daniel Forsberg \inst{1}\orcidID{0000-0001-7812-4167}}
\authorrunning{A. Berg, D. Forsberg}
%
\institute{Department of Computer and Systems Sciences, Stockholm University, Stockholm, Sweden}
\maketitle              
\begin{abstract}

DNS is a distributed, fault tolerant system that avoids a single point of failure. As such it is an integral part of the internet as we use it today and hence deemed a safe protocol which is let through firewalls and proxies with no or little checks. This can be exploited by malicious agents. Network forensics is effective but struggles due to size of data and manual labour. 
This paper explores to what extent predictive models can be used to predict network traffic, what protocols are tunneled in the DNS protocol and more specifically whether the predictive performance is enhanced when analyzing DNS-queries and responses together and which feature set that can be used for DNS-tunneled network prediction. The tested protocols are SSH, SFTP and Telnet and the machine learning models used are Multi Layered Perceptron and Random Forests. To train the models we extract the IP Packet length, Name length and Name entropy of both the queries and responses in the DNS traffic. With an experimental research strategy it is empirically shown that the performance of the models increases when training the models on the query and respose pairs rather than using only queries or responses. The accuracy of the models is $>83\%$ and reduction in data size when features are extracted is roughly $95\%$. Our results provides evidence that machine learning is a valuable tool in detecting network protocols in a DNS tunnel and that only an small subset of network traffic is needed to detect this anomaly.

%

\end{abstract}

\section{Introduction}

\pagestyle{plain}
\setcounter{page}{1}
\pagenumbering{arabic}

The Internet Protocol (IP) \cite{postel1981rfc} can be used to send IP-packages between computers in an interconnected network. These IP-packages can be freely inspected by anyone handling the package, and be used to handle data delivery needed for a variety of different protocols, such as the Domain Name System (DNS). DNS is today an integral and important part of the Internet and it is the protocol that resolves domain names. If DNS did not exist as a translator between humans and computers, for every website a user wanted to visit they would have to know the IP-address of the computer that hosts that particular website \cite{rfc1035}. Easily recognizable names, such as www.wikipedia.org, is easier to remember for humans than an IP-address, such as 207.142.131.234. The DNS system is the backbone of the World Wide Web (WWW) as we use it today \cite{berners1994rfc}.

As such, the DNS system is a distributed, fault tolerant system that avoids a single point of failure. A client iterates queries to DNS servers until it gets a response for the queried endpoint, for instance a web server. If a DNS server does not have the IP-address for a queried endpoint it simply responds with another DNS server that might have the correct IP-address.



The practice of \emph{tunneling} network traffic means that the data that is to be transported over a network is chopped up into smaller pieces called Protocol Data Unit (PDU), \emph{encapsulated} with a header for the specific protocol. Depending on how many protocols are involved in transporting the data the process of encapsulating a PDU can be done several times before it is transported to another computer over the network. The receiving computer then puts the data back together by decapsulating the data according to the specified protocol(s) used.

As described here tunneling of network traffic is the usual way of transporting data across a network. Protocols such as the Secure Shell protocol (SSH) allows communication between computers using encryption techniques over an insecure network \cite{ylonen2005secure}. By tunneling the activities on the remote computer through the SSH protocol a remote user can connect to, perform actions and see the result of those actions as if they were performed on the users local computer.

%

However, the same technique, to tunnel network traffic, can be used to hide malicious traffic in a standardized protocol by repackaging the malicious traffic to look like normal traffic. By repackaging for instance a SSH connection to look like DNS traffic and sending that traffic over the network any software or computer looking at that traffic will see it as DNS traffic. This is done in order to circumvent firewalls or proxies to be able to compromise networks or individual computers \cite{DUSI200981}.


The type of malicious traffic that exploits standardized protocols, like DNS, can hide their traffic inside these protocols since they are usually let through the outer perimeter of any LAN guarded by a firewall with no, or few, checks as described by Xu, Butler, Saha, and Yao \cite{xu2013dns}. The normal checks performed on network traffic by firewalls and proxies before allowing access to resources are not performed on DNS since it is deemed a safe protocol only relaying DNS lookups to local computers \cite{born2010detecting}.

This type of stealthy network traffic can focus on hiding traffic between malware and their servers, which is common among botnets, as well as sending sensitive data to malicious agents or mapping out an organizations LAN \cite{xu2013dns}. Network forensics in this area, that is work being conducted in order to detect malicious network traffic, is mostly focusing on reactive measures which entails storing large amounts of network traffic for time consuming \emph{manual analysis}, requiring \emph{skill} and \emph{storage space} which could implicate \emph{privacy problems} \cite[p. 19-21]{davidof2012network}. 

Machine learning is the use of algorithms to build, or train, mathematical models based on a sample set of data in order to be able to make predictions or decisions without being explicitly programmed for that task. Machine learning algorithms are used in a wide variety of applications such as spam email filtering or face recognition in pictures. \cite[p. 1-2]{alpaydin2009introduction} 

\subsection{Problem}

Automatically detecting tunneled, malicious traffic has been tried and the results shows promise. Born and Gustafson \cite{born2010detecting} matched Zipfian distribution of the English language to the character frequency of the network traffic to detect DNS tunnels. Homem and Papapetrou \cite{homem2017harnessing} and Homem, Papapetrou, and Dosis \cite{homem2017entropy} have used machine learning to find what protocols where tunneled in through the DNS protocol with an accuracy of $\geq 90\%$ detection rate for the tested protocols. Despite the efforts in this research area, detecting tunneled traffic in a standardized protocol, using machine learning models is a somewhat neglected area apart from~\cite{homem2017harnessing} which has focused on four protocols; HTTP, FTP, HTTPS, POP3 and using four predictive models: K-NN, Decision Trees, SVM and Multinomial Neural Networks. Also, the datasets that has been utilized in previous research has been rather small limiting the research to a rather narrow spectrum of traffic type.

Although the efforts of Homem and Papapetrou \cite{homem2017harnessing} shows promise, as mentioned above it uses only a small dataset with just a few protocols which limits them to a small set of features such as \emph{DNS Query Name Entropy, DNS Query Name Length} and the \emph{IP Packet Length} for data mining, only using information from the DNS-queries and neglecting the responses. Based on the above there exist a knowledge gap in the area of detecting tunneled, malicious network traffic with predictive models posing a need for a comprehensive experiment where a broad dataset in terms of traffic type is utilized to gain new insights to what features are important, especially since Homem and Papapetrou \cite{homem2017harnessing} did not experiment with protocols such as SSH, SFTP or Telnet among others.

\subsection{Contributions}

The main contributions of this paper are: 

\begin{description}

\item[New features and combinations]
%
While earlier research such as Homem and Papapetrou \cite{homem2017harnessing} and Nadler et al. \cite{nadler2019detection} has shown the effectiveness of their proposed methods, they have not done any comparisons between different feature-sets. They don't make any attempts to analyze the inherit query-response structure from the DNS protocol. This paper compares the different model's and feature-sets accuracy and show that the predictions are better when analyzing the query and the response packets together.

\item[Next level network forensics]
We present a basic Network Forensic Tool\footnote{\url{https://gitlab.com/vt19/networkforensictool}} that can be used as a post mortem tool for Network Forensics to detect tunneled DNS-traffic and is able to analyze DNS-queries and responses as pairs. This allows feature-sets that corresponds to the inherit query-response structure of the DNS protocol. Adding to the flexibility in the feature-set construction, setting it apart from tools such as the ones presented by Homem and Papapetrou \cite{homem2017harnessing}.


\end{description}
%
%
%
%

\section{Related work}
Below we summarize related work in DNS tunnels, Network forensics and Machine Learning.


Botnets i.e.\ sets of computers that are infected by a specific malware allowing them to be remote controlled are a major security issue on the Internet. Since the Command and Control (C\&C) is a defining characteristic of botnets, techniques have been developed to detect bot infections by identifying the C\&C network traffic \cite{dietrish2011dnscc}. Botnet operators have tried different structures to improve the resilience of their C\&C. The botnet Feederbot have adapted by abusing the DNS protocol and uses DNS tunneling for its C\&C \cite{dietrish2011dnscc}. 

The DNS protocol provides a distributed infrastructure for storing, updating and disseminating data that conveniently fits the need for a C\&C system. Not only does it provide a means of communication between computers, but also mechanisms for naming, locating, distributing and caching resources with fault tolerance. \cite{xu2013dns}

\begin{table}
  \begin{tabular}{l r p{0.7\textwidth}} \toprule
    \multicolumn{3}{c}{DNS tunneling data holders} \\ \cmidrule{1-3}
    Record Name & Max Size & Description \\ \midrule
    CNAME & 255 & a domain name \\
    MX & 255  &  a 16 bit preference value (lower is better) followed by a host name willing to act as a mail exchange for the owner domain. \\
    TXT & 255 & used to hold descriptive text \\
    EDNS0 & &an extension that allows a payload bigger than the original 512-byte maximum.\cite{rfc6891} When a capable server or client encounters this field, it can decode the packets, allowing several improvements to the basic DNS protocol. These features include larger UDP packet size, a list of attribute value pairs, and several extra bytes for commonly used flags. \\
    \bottomrule
  \end{tabular}
  \caption{DNS records suitable for use in DNS tunneling \cite{xu2013dns}. Note that without EDNS0 the entire response of the query, which includes the query and response, must be smaller than 512 bytes.\cite{rfc1034,rfc1035}}\label{tab:dnsrecords}
\end{table}

The DNS system allows a name server administrator to associate different types of data with either a fully qualified domain name or an IP address. To send a message to a bot, an adversary can store data in any one of these types of records described in table~\ref{tab:dnsrecords}.
These records can be used by the clients to establish a two-way transfer of data by encoding queries to the server, which then can send data back encoded in the records. \cite{xu2013dns}


The purpose of Network Forensics is to monitor and analyze network traffic for information gathering. This information can be used for legal or intrusion detection purposes. Network traffic as such is volatile and dynamic meaning investigations in this area needs to be pro-active \cite{palmer2001road}.

There is only so much information stored in log files and if an attacker has erased the logs there will be a hard time retracing the attackers actions and entry point into the system. This means that forensics in this area either needs to collect and store real-time data for later analysis or try to analyze the data as it flows through the network. The collection of data can cause privacy concerns in itself as questions such as \emph{Who} will collect it, \emph{When} or how often it should be collected and \emph{What} is to be collected \cite[p. 29]{palmer2001road}. This collected traffic is then analyzed by a forensic analyst, often manually which is time consuming, error-prone and can constitute serious privacy issues \cite[p. 19-21]{davidof2012network}


The scalability of a Network Forensics Tool where large amounts of data needs to be stored centrally on a forensic server is an critical issue that is discussed by Khan, Gani, Wahab, Shiraz, and Ahmad \cite{KHAN2016214}. Any tools used in network forensics needs to improve on the manual labour put down in order to be a viable option and considering the amount of network traffic that is generated today needs to scale to handle large amounts of data.
Considering the amount of data that is transmitted in modern networks it is bound to be a lot of noise, i.e. data or information that is not malicious and not of real interest to an analyst. This data would also most likely contain sensitive information such as passwords or even financial records further increasing the privacy issues, both personal and organizational. \cite[p. 231]{KHAN2016214}

The research of Liao, Tian, and Wang \cite[p. 1883]{liao2009network} using fuzzy set \cite{zadeh1965fuzzy} logic on extracted features of the collected network data show that only parts of the data is actually needed to detect malicious traffic. This is in accordance with Homem and Papapetrou \cite{homem2017harnessing} findings and would address at least some of the privacy issues stated earlier. However, Liao et al. \cite{liao2009network} results also show that the computing costs of extracting new fuzzy rules is large which could lead to new attacks not being detected in time. This is still better than the manual analysis performed on collected data and the use of machine learning could act as a sieve on large amounts of data for an analyst to only look at the parts of traffic that are likelier of being significant for forensic research.

To classify the data Homem and Papapetrou \cite{homem2017harnessing} got their best results using Multinomial Neural Networks where they used a Multi-Layer Perceptron (MLP) to classify the dataset. The MLP is trained by a already classified dataset and are presented to a feed-forward back-propagation neural network, with every feature having a input node and every classifier a respective output node \cite{nanopoulos01featurebasedclassification}. Each node has a response $f(w^tx)$ where $x$ is the vector of output activations from the preceding layer, $w$ is a vector of weights and $f$ is a bounded nondecreasing nonlinear function, i.e. the sigmoid \cite[p. 31]{reed1999nss}.



Other interesting potential techniques include random forest. Random forest are a combination of tree predictors such that each tree depends on the values of a random vector sampled independently with the same distribution for all trees in the forest. Where the classification function simply output the most popular answer among the trees when given a data-point. The Random forests can be constructed by giving every tree a randomly selected input set and training it using random tree specific choice of features. Such random forests produce good results in classification. \cite{breiman2001random}

The use of entropy measurement is an interesting feature that can be used quite effectively to classify DNS-traffic as DNS-tunneled. \cite{homem2017entropy}. The entropy can be measured by Shannon's entropy formula, $H(X) = -\sum_{k=1}^{n}p_k*\ln_e p_k$  where X is constructed by an alphabet $\{x_1, x_2, ..., x_n\}$ and $p_k$ is the relative frequency for the respective characters. \cite[pp. 59-61]{arnd2001information}

\subsection{Recent research}

Although there is recent research in the area exploring different approaches to detecting and classifying protocols within tunnels of other protocols such as Dusi et al. \cite{DUSI200981} who try to distinguish normal HTTP and SSH traffic from malicious tunnels. Alshammari and Zincir-Heywood \cite{alshammari2011can} used classifiers such as Adaboost to try and detect the actual application traffic such as X11 in an SSH-tunnel.

Research not focusing on tunneling per se but at identification of obfuscated protocols, that is protocols that makes it harder to identify them by i.e.\ randomizing their payload, is notably Hjelmvik and John \cite{hjelmvik2010breaking}. In their research they could correctly identify $\geq90\%$ of a large part of the tested protocols with the Kullback-Leibler divergence \cite{kullback1951information} on subsets of extracted features from the collected network traffic data.

In their study examining how well the entropy of DNS-tunneled traffic could be predicted using a simple \emph{MeanDiff} prediction Homem et al. \cite{homem2017entropy} were able to make quite accurate classifications. In a later study Homem and Papapetrou \cite{homem2017harnessing} expanded upon the previous study, studying the accuracy of four common machine learning techniques to classify different traffic. They used k-Nearest-Neighbour (k-NN), Decision Trees, Support Vector Machines (SVMs) and Neural Networks, upon a dataset of HTTP, FTP, HTTPS and POP3 through a DNS-tunnel reduced to three factors. These specific factors were \emph{DNS Query Name Entropy}, \emph{DNS Query Name Length}, and the \emph{IP Packet Length}. With this approach they were able to reach at least 90\% accuracy classifying the traffic. \cite{homem2017harnessing}

\section{Network traffic classification Work flow}

According to sound, robust and transparent research methodology principles, a strategy is necessary when conducting research. This is important as it formalizes the way in which the research itself is conducted and therefore gives a certain amount of validity and reliability as it makes the research reproducible for others to either expand on or control earlier research results \cite{golafshani2003understanding}.

In our experiments the following work flow was followed:

\begin{enumerate}
    \item Record DNS traffic with a tool such as tcpdump
    \item Extract the desired features with a Network Forensic Tool
    \item Apply a Machine Learning model on the extracted features 
    \item Evaluate the results
\end{enumerate}

\subsection{Evaluation methodology}\label{sec:evalmetod}

The findings of this paper is evaluated using the metrics Accuracy (ACC), Precision (P) and Recall (R). These metrics uses the True Positives (TP), True Negatives (TN), False Positives (FP) and False Negatives (FN) in calculating how well the predictive models perform. 


A 20-fold cross-validation was used to evaluate the algorithms consisting of the models feature-sets and compared against the results of the Multinomial Neural Networks using the \cite{homem2017harnessing} featureset. The 20-fold cross-validation technique consists in splitting a dataset in 20 independent subsets: in turn all but one of these subsets are used to train a classifier, while the remaining one is used to evaluate the generalization error \cite{anguita2009k}. We will use the models and feature-sets presented in table~\ref{tab:models} and when calculating the entropy we will consider the single bytes the characters in our alphabet.

\begin{table}
  \begin{tabular}{l p{0.4\textwidth}@{\hspace{8pt}}p{0.4\textwidth}} \toprule
   & \multicolumn{2}{c}{Model} \\ \cmidrule{2-3}
Feature set & Multinomial Neural Network & Random Forest \\ \midrule
Query & Name Entropy, Name Length and IP Packet Length of queries only & Name Entropy, Name Length and IP Packet Length of queries only\\
Full & Name Entropy, Name Length and IP Packet Length of queries and responses together & Name Entropy, Name Length and IP Packet Length of queries and responses together\\
Response & Name Entropy, Name Length and IP Packet Length of responses only & Name Entropy, Name Length and IP Packet Length of responses only\\
\bottomrule
  \end{tabular}
  \caption{The classification algorithms consisting of model and feature set choice used in this paper}\label{tab:models}
\end{table}

The statistical significance of the results is ensured with a Friedman test \cite{friedman1940comparison}. This tests whether the classification is significantly different in accuracy between the algorithms. The null hypothesis is that there are no differences in classification accuracy between the algorithms. If the null hypothesis is rejected the Friedman test will be followed by a post-hoc test evaluating whether the classification of the algorithms are different from the algorithm using Multinomial Neural Networks and the Homem and Papapetrou \cite{homem2017harnessing} feature set. The post-hoc test will use the Holm step-down procedure \cite{holm1979simple}. 

In the Friedman test the algorithms are assigned ranks for each data-set, the best is assigned rank 1, the second rank 2..., ties are assigned the average of the ranks. We assign the ranks after the accuracy of the different algorithms for each fold. The Friedman statistic,
\begin{equation}
\chi^2_F = \frac{12N}{k(k+1)}\left(\sum_jR^2_j - \frac{k(k+1)^2}{4} \right)
\end{equation}
where $R_j = \frac{1}{N}\sum_ir_i^j$ and $r_i^j$ is the rank of the $i$'th of the $k$ algorithms and the $j$'th of the $N$ datasets, is used to calculate $F_F$ \cite{friedman1940comparison}.
\begin{equation}
F_F = \frac{(N-1)\chi_F^2}{N(k-1)-\chi_F^2}
\end{equation}
$F_F$ is distributed according to the $F$-distribution with $k-1$ and $(k-1)(N-1)$ degrees of freedom \cite{iman1980approximations}. The $F_F$ is compared to the critical value of this $F$-distribution with $\alpha=0.05$. If $F_F$ is bigger than the critical value the null hypothesis is rejected. The null hypothesis is that there are no difference of accuracy between the algorithms. The post-hoc test checks whether there is a significant difference the chosen algorithms. 
The test statistic for comparing an algorithm to our set algorithm, here given index $0$ is
\begin{equation}
z_i = (R_i-R_0)\left/\sqrt{\frac{k(k+1)}{6N}}\right. 
\end{equation}
The $z_i$-values are used to find their respective $p_i$-values whether the $z_i$ follows the normal distribution with mean 0 and standard deviation (SE) $\sqrt{\frac{k(k+1)}{6N}}$. In accordance to the Holm's step-down test the $p_i$-values are sorted and the indices are reassigned such that $p_i \leq p_j$ for all $i,j\in\mathbb{Z}^{\geq 1}, i < j$. These $p_i$ values are checked algorithm for algorithm in the order of the indices whether the $p_i$-value is lower than the $\alpha/(k-i)$, for the chosen significance level $\alpha$. If $p_i < \alpha/(k-1)$ then the null hypothesis is rejected, ie there is a significant difference between the $i$ and the $0$ algorithm. This is continued until a $p_i$ is unable to reject the null hypothesis, all following null hypothesis are then accepted. \cite{holm1979simple,demvsar2006statistical}




\subsection{Data sets}
The DNS-tunneled data was generated in a controlled environment.\footnote{\url{https://s3.eu-central-1.wasabisys.com/dns-tunneling/dns_tunnel_sftp.pcapng}}\footnote{\url{https://s3.eu-central-1.wasabisys.com/dns-tunneling/dns_tunnel_ssh.pcapng}}\footnote{\url{https://s3.eu-central-1.wasabisys.com/dns-tunneling/dns_tunnel_telnet.pcapng}} This is since abusing the DNS-protocol can be considered malicious or even illegal. The non-DNS-tunneled data is a dataset collected and provided by CAIDA (Center for Applied Internet Data Analysis) at the University of California San Diego (UCSD) for research purposes. This data is used in accordance with their \emph{CAIDA Acceptable Use Agreement} \cite{caida2014qurterly,caida2019topology}. The dataset consists of approximately one GiB of DNS traffic in pcap files compressed by GNU Zip. The data will be classified according to what type of protocol it DNS-tunnels and a class for all non-tunneled DNS-traffic. From the datasets the specified features is then extracted and used to train the models.

\section{Results}
\label{cha:result}

For this study we created our own dataset of DNS-tunneled traffic. To create the dataset we set up an authoritative DNS server for the domain \url{dnshax.se} running the DNS-tunneling software \emph{DNSCat2}\footnote{\url{https://github.com/iagox86/dnscat2}}, a set of virtual computers, hosted on one of our own computers, to control through the DNS-tunnel. On the host computer running the virtual computers we recorded the network traffic on the DNS-protocol with the tool \emph{tcpdump}\footnote{\url{https://www.tcpdump.org/}} for the virtual computers specific IP-addresses.

The total size of the dataset is $>4$Gb worth of recorded .pcap files. When the feature set had been extracted from the .pcap files into comma separated text files the total size was $\approx 229$Mb indicating a $95\%$ reduction in space requirements. This could have been reduced even further by storing the extracted features in a binary format.
Our network forensics tool were coded in python, using the \emph{scapy}\footnote{\url{https://scapy.net/}} framework for the feature extraction; for the models we used implementations from the \emph{scikit-learn}\footnote{\url{https://scikit-learn.org}} framework with the default settings for the respective models. 

The extracted features we used for this experiment had $1\,101\,594$ features, that is responses and queries, per tested protocol totaling $4\,406\,376$ features in all the feature sets. These were all crunched through the respective models with the results detailed below.


\subsection{Results on query and response pair}%
\label{sec:results_on_query_and_response_pair}
\begin{table}[h]
	\begin{tabular}{l *{6}{@{\hspace{6pt}}c}} \toprule 
		Feature sets & \multicolumn{6}{c}{Full }\\ 
\cmidrule(l){2-7}
		Models & \multicolumn{3}{c}{Neural Network }& \multicolumn{3}{c}{Random Forest }\\ \midrule 
		\diagbox[width=2.5cm, height=2.5cm]{Protocols}{\rotatebox{90}{Metrics (\%)}} & \rotatebox[origin=d]{90}{\bf Precision}& \rotatebox[origin=d]{90}{\bf Recall}& \rotatebox[origin=d]{90}{\bf Accuracy}& \rotatebox[origin=d]{90}{\bf Precision}& \rotatebox[origin=d]{90}{\bf Recall}& \rotatebox[origin=d]{90}{\bf Accuracy}\\ 
		\bf DNS& \small 1.000 & \small 1.000 & \multirow{4}{*}{\small0.830} & \small 1.000 & \small 1.000 & \multirow{4}{*}{\small0.839} \\ 
		\bf SFTP& \small 1.000 & \small 0.999 & & \small 1.000 & \small 0.999 & \\ 
		\bf SSH& \small 0.599 & \small 0.977 & & \small 0.657 & \small 0.747 & \\ 
		\bf TELNET& \small 0.936 & \small 0.345 & & \small 0.706 & \small 0.610 & \\ 
		\bottomrule 
	\end{tabular} 
  \caption{The evaluation metrics of the models using the full feature set presented in table~\ref{tab:models}}\label{tab:full_metrics} 
\end{table}

As can be seen in table \ref{tab:full_metrics} none of the models have problem distinguishing normal DNS traffic from malicious traffic as the Precision and Recall are both $1.000$. Among the tunneled protocols SFTP was the one that was easiest to spot for both models when using the combination of query and response pairs with precision at $1.000$ and recall at $0.999$ for both models as can be seen in table \ref{tab:full_metrics}. 

Both models have problem distinguishing between tunneled SSH and Telnet traffic when using both queries and responses in the feature set as can be seen in table \ref{tab:full_metrics}. The neural networks recall is really low for Telnet at $0.345$ compared to the random forest at $0.610$. 

Accuracy is a respectable $0.830$ for the Neural Network and $0.839$ for the Random Forest for the entire feature set over all the protocols.

\subsection{Results on query feature set}%
\label{sec:results_on_query_feature_set}
\begin{table}[h]
	\begin{tabular}{l *{6}{@{\hspace{6pt}}c}} \toprule 
		Feature sets & \multicolumn{6}{c}{Query }\\ 
\cmidrule(l){2-7}
		Models & \multicolumn{3}{c}{Neural Network }& \multicolumn{3}{c}{Random Forest }\\ \midrule 
		\diagbox[width=2.5cm, height=2.5cm]{Protocols}{\rotatebox{90}{Metrics (\%)}} & \rotatebox[origin=d]{90}{\bf Precision}& \rotatebox[origin=d]{90}{\bf Recall}& \rotatebox[origin=d]{90}{\bf Accuracy}& \rotatebox[origin=d]{90}{\bf Precision}& \rotatebox[origin=d]{90}{\bf Recall}& \rotatebox[origin=d]{90}{\bf Accuracy}\\ 
		\bf DNS& \small 1.000 & \small 1.000 & \multirow{4}{*}{\small0.819} & \small 1.000 & \small 1.000 & \multirow{4}{*}{\small0.817} \\ 
		\bf SFTP& \small 0.972 & \small 0.997 & & \small 0.971 & \small 0.998 & \\ 
		\bf SSH& \small 0.598 & \small 0.941 & & \small 0.629 & \small 0.730 & \\ 
		\bf TELNET& \small 0.850 & \small 0.340 & & \small 0.667 & \small 0.540 & \\ 
		\bottomrule 
	\end{tabular} 
  \caption{The evaluation metrics of the models using the query feature set presented in table~\ref{tab:models}}\label{tab:query_metrics} 
\end{table}

When only using the queries in training the models they still have no problem spotting the normal DNS traffic as it is still at $1.000$ for Precision and Recall for both models as can be seen in table \ref{tab:query_metrics}. Both models also produce similar results for SFTP with only minor differences between them when only using the query feature set with Precision $>0.97$ and Recall close to $1.000$.

When only using the query feature set both models still have problems differentiating between tunneled SSH and Telnet traffic as can be seen in table \ref{tab:query_metrics}. Both models have a fairly low Recall for Telnet at $0.340$ for the Neural Network and $0.540$ got the Random Forest.

The accuracy for both models on only the query feature set is slightly lower than the results for both queries and responses at $0.819$ for the Neural Network and $0.817$ for the Random Forest.

\subsection{Results on response feature set}%
\label{sec:results_on_response_feature_set}
\begin{table}[h]
	\begin{tabular}{l *{6}{@{\hspace{6pt}}c}} \toprule 
		Feature sets & \multicolumn{6}{c}{Response }\\ 
\cmidrule(l){2-7}
		Models & \multicolumn{3}{c}{Neural Network }& \multicolumn{3}{c}{Random Forest }\\ \midrule 
		\diagbox[width=2.5cm, height=2.5cm]{Protocols}{\rotatebox{90}{Metrics (\%)}} & \rotatebox[origin=d]{90}{\bf Precision}& \rotatebox[origin=d]{90}{\bf Recall}& \rotatebox[origin=d]{90}{\bf Accuracy}& \rotatebox[origin=d]{90}{\bf Precision}& \rotatebox[origin=d]{90}{\bf Recall}& \rotatebox[origin=d]{90}{\bf Accuracy}\\ 
		\bf DNS& \small 0.998 & \small 0.994 & \multirow{4}{*}{\small0.585} & \small 0.999 & \small 0.999 & \multirow{4}{*}{\small0.821} \\ 
		\bf SFTP& \small 0.404 & \small 0.657 & & \small 1.000 & \small 0.999 & \\ 
		\bf SSH& \small 0.415 & \small 0.302 & & \small 0.634 & \small 0.677 & \\ 
		\bf TELNET& \small 0.593 & \small 0.385 & & \small 0.651 & \small 0.608 & \\ 
		\bottomrule 
	\end{tabular} 
  \caption{The evaluation metrics of the models using the response feature set presented in table~\ref{tab:models}}\label{tab:response_metrics} 
\end{table}

When looking in table \ref{tab:response_metrics} we can see that both models are just shy of $1.000$ in precision and recall for normal DNS traffic. What really stands out though compared to the results in sections \ref{sec:results_on_query_and_response_pair} and \ref{sec:results_on_query_feature_set} is that the Neural Network performs really bad for spotting SFTP traffic at a precision and recall of $0.404$ and $0.657$ respectively compared to the Random Forest at $1.000$ and $0.999$.

For the other protocols the Neural Network performs worse than the Random Forest although not as badly as for SFTP as can be seen in table \ref{tab:response_metrics}. The Neural Networks lack of performance when only using the responses as the feature set lands the models on an accuracy of $0.585$ and $0.821$ respectively.

%
%
\begin{table}
	\begin{tabular}{l l c c c } \toprule 
\multicolumn{5}{l}{$\sum_{i}R_i^2$ = 89.089, $N$ = 20, $k$ = 6, $\chi_F^2$ = 89.079} \\ 
\multicolumn{5}{l}{$F_F$ = 154.970, $df_1$ = 5, $df_2$ = 95, $\alpha$ = 0.05, $F_F$ critical value = 2.310} \\
\midrule
  i & Algorithm & $z = \left(R_i - R_0\right)/SE$ & $p$ & $\alpha/(k-i)$ \\ \midrule
1 & Response, Neural Network & 7.429& 0.000 & 0.01\\
2 & Full, Random Forest & -6.857& 0.000 & 0.013\\
3 & Query, Random Forest & 3.786& 0.000 & 0.017\\
4 & Full, Neural Network & -3.286& 0.000 & 0.025\\
5 & Response, Random Forest & 0.643& 0.211 & 0.05\\
		\bottomrule 
	\end{tabular} 
	\caption{The results of the Friedman test and the post-hoc Holm step-down procedure}\label{tab:friedman} 
\end{table}

\subsection{Significance analysis}%
\label{sec:significance_analysis}
The results from the Friedman test and the post-hoc analysis are presented in table \ref{tab:friedman}.
The $F$-distributed $F_F$ statistic, derived from the Friedman statistic ($\chi^2_F$), is bigger than the critical value and the first null hypothesis is therefore rejected, ie there exists a significant difference in accuracy between the different classification algorithms. 
Further the post-hoc analysis were unable to reject the null hypothesis that the classification algorithm using the responses feature set and Random Forest model are not significantly different. All the other algorithms were shown to be significantly different from the algorithm using Multinomial Neural Network with the Homem and Papapetrou \cite{homem2017harnessing} feature-set.

\section{Conclusions}%
\label{sec:conclusion}

In our results we have shown that predictive models can predict network traffic as can be seen in tables \ref{tab:full_metrics}, \ref{tab:query_metrics}, \ref{tab:response_metrics} and \ref{tab:friedman}. Although there are differences between the different models and protocols our results clearly show that predictive models can, and do, predict tunneled network traffic.


We have also shown that the models do perform better when combining the queries with the corresponding responses instead of only using queries or responses when trying to predict network traffic.


The amount of data needed to store for analyzing network traffic is reduced by $95\%$ and could be reduced even more, mitigating several storage and privacy issues.

Also, when using only queries or responses, the models have variable performance, i.e. one model is better at predicting traffic than the other on a specific feature set, suggesting the model is the determinant, not the feature set.

 \bibliographystyle{splncs04}
 \bibliography{biblio}

\end{document}